\documentclass{PoS}
\usepackage{graphicx}

\PoS{PoS(LAT2005)154}

\title{Remarks on the Maximum Entropy Method applied to finite temperature
  lattice QCD }

\ShortTitle{Remarks on the Maximum Entropy Method applied to finite
  temperature lattice QCD }

\author{\speaker{Takashi Umeda}\thanks{A footnote may follow.}\\
        Brookhaven National Laboratory,
        Upton, New York, 11973, USA\\
        E-mail: \email{tumeda@quark.phy.bnl.gov}}

\author{Hideo Matsufuru\\
        High Energy Accelerator Research Organization (KEK),
        Tsukuba 305-0801, Japan\\
        E-mail: \email{hideo.matsufuru@kek.jp}}

\abstract{We make remarks on the Maximum Entropy Method (MEM) for
studies of the spectral function of hadronic correlators in finite
temperature lattice QCD. 
We discuss the virtues and subtlety of MEM in the cases that one does
not have enough number of data points such as at finite temperature.
Taking these points into account, we suggest several tests which one
should examine to keep the reliability for the results, and also apply
them using mock and lattice QCD data.}

\FullConference{XXIIIrd International Symposium on Lattice Field Theory\\
		 25-30 July 2005\\
		 Trinity College, Dublin, Ireland}

\begin{document}

\section{ Introduction }

In the lattice QCD, most properties of hadrons are extracted from the 
hadronic correlation functions.
The spectral function (SPF) has particular importance, since
it may contain information beyond the stable ground state and a few 
excited states which can be extracted by standard fitting techniques.
Recent development of analysis techniques such as the maximum entropy
method (MEM) \cite{nakahara99} have enabled direct extraction of
SPFs from numerical data of lattice QCD simulation.
At zero temperature, MEM has been successfully reproduced
correct features of the SPFs \cite{nakahara99,yamazaki01}.
%

At finite temperature, we can calculate the SPF
from the thermal green functions in principle using the
same procedure as at zero temperature \cite{abrikosov59, taro01}.
In particular, charmonium states have drawn much attention,
since they probe the QCD plasma state
through the changes of their properties
\cite{hashimoto86, Matsui:1986dk},
and hence are potential signal of the formation of 
quark gluon plasma in the heavy ion collision
experiments \cite{na50}.
%
Several groups have studied the SPF of
charmonium in finite temperature lattice QCD using MEM
\cite{umeda02,asakawa03, datta03}
and their results indicate persistent $J/\psi$ state even above
$T_c$. 
MEM has also been extensively applied to various 
areas of lattice field theories \cite{MEMothers}. 

While MEM is a powerful tool to extract SPF,
it has intrinsic subtlety when applied to lattice QCD data
of correlators.
In this paper, we point out how each ingredient of MEM analysis
causes such subtlety, focusing on an application to the
correlators at finite temperature.
In the next section, we consider general problems of MEM,
and then in Sect.~\ref{sect:ft} describe particular problem
at finite temperature caused by short extent in the temporal
direction.
Details of these analysis were presented in Ref.~\cite{umeda02}.

%

\section{ Maximum entropy method }
\label{sect:mem}
\subsection{ Outline of MEM }
\label{sect:omem}

First we briefly summarize the outline of MEM basically following
Ref.~\cite{nakahara99}, which reviews in detail MEM applied to data
of lattice QCD simulation.
We obtain the SPF, $A(\omega)$, from the 
given lattice result for the correlator, $C(t)$, 
by solving the inverse problem, 
\begin{equation}
C(t) = \int_0^{\infty} d\omega K(t,\omega) A(\omega),
\end{equation}
where the (continuum type) kernel $K(t,\omega)$ is given by
\begin{equation}
 K(t,\omega)=\frac{ e^{-\omega t}+e^{-\omega(N_t-t)} }
                     { 1 - e^{-N_t\omega} }.
\label{eq:kernel}
\end{equation}
To extract the SPF $A(\omega)$, MEM maximizes a
functional $Q(A;\alpha) = \alpha S[A] - L[A]$.
$L[A]$ is the usual likelihood function,
and minimized in the standard $\chi^2$ fit.
The Shannon-Jaynes entropy $S[A]$ is defined as
\begin{equation}
 S[A] = \int_0^\infty d\omega
   \left[ A(\omega)-m(\omega)-A(\omega)
           \log \left( \frac{A(\omega)}{m(\omega)}\right) \right].
\end{equation}
The function $m(\omega)$ is called the default model function, 
and should be given as a plausible form of $A(\omega)$.
At the last stage of calculation the parameter $\alpha$ can be
integrated out by a weighted average of prior probability for $\alpha$.

\subsection{ Singular Value Decomposition }
\label{sect:svd}

In the maximization step of $Q(A;\alpha)$ the singular value
decomposition of the kernel $K(t,\omega)$ is usually used
\cite{nakahara99}.\footnote{
Analysis of MEM without singular value decomposition was examined
in Ref.~\cite{Fiebig:2002sp}.}
%
Then the SPF is represented as a linear
combination of the eigenfunctions of $K(t,\omega)$:
\begin{eqnarray}
 A(\omega)=m(\omega)
\exp\left\{ \sum_{i=1}^{N_s} b_i u_i(\omega) \right\},
\label{eq:spec_svd}
\end{eqnarray}
where $N_s$ is the number of eigenfunctions, $b_i$ are parameters, and 
$u_i(\omega)$ the eigenfunction of the kernel $K(t,\omega)$.
The number of degrees of freedom of $A(\omega)$ is accordingly
reduced to the number of data points of the correlator.
Although $b_i$
can in principle be determined uniquely from the data without
introducing an entropy term, the small eigenvalues of $K(t,\omega)$
lead to a singular behavior of the SPF;
hence truncation of the terms is practically required,
i.e. $N_s$ may be less than the number of data points
\cite{taro98}.  
In MEM, the entropy term stabilizes the problem and
guarantees an unique solution for the coefficients of the eigenfunctions
\cite{nakahara99}.

\begin{figure}[tb]
\center{
\includegraphics[width=50mm]{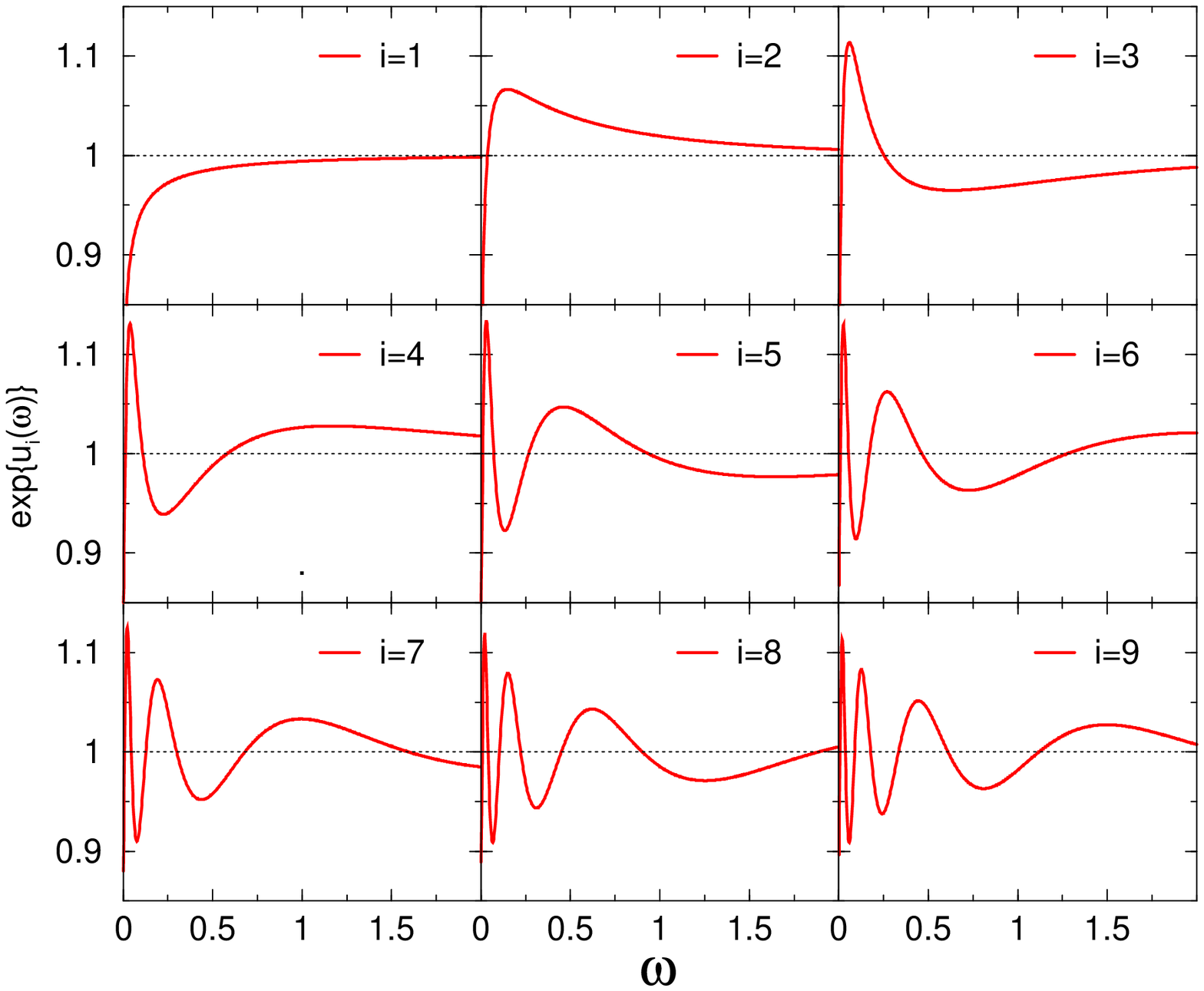}
\includegraphics[width=50mm]{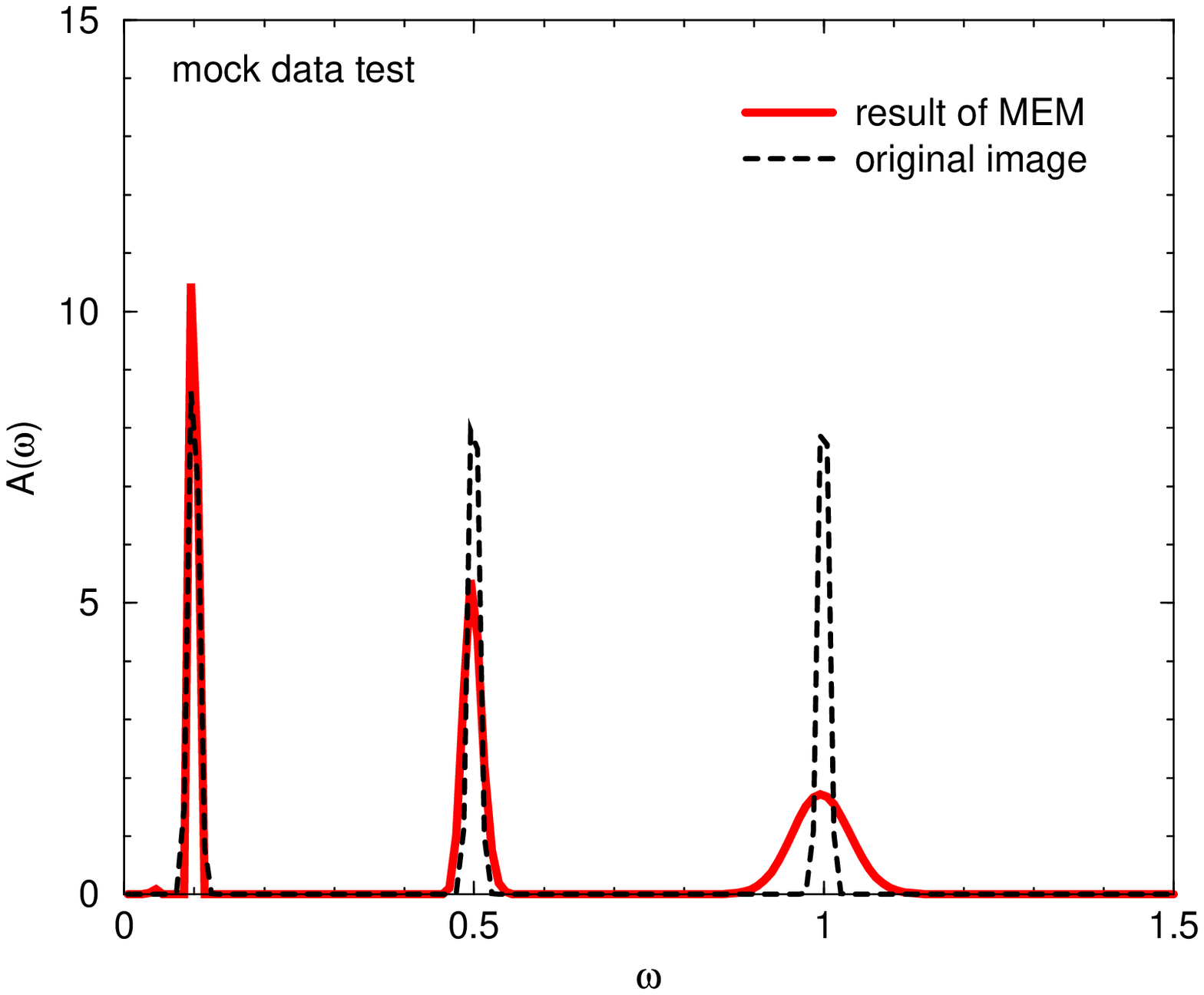}
\includegraphics[width=44mm]{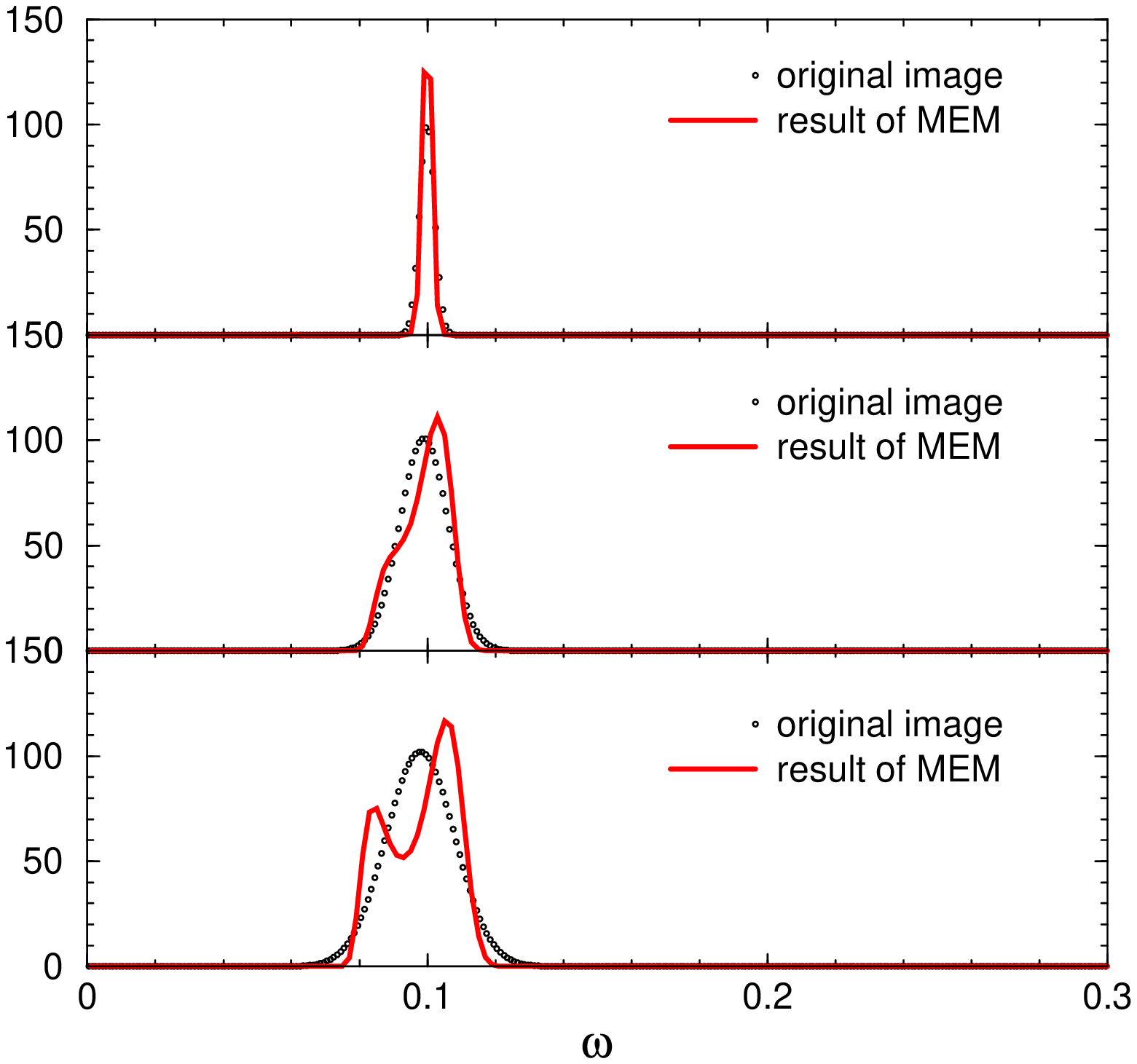}}
\vspace{-0mm}
\caption{Left panel: Samples of eigenfunction for the kernel,
Eq.~(2.2). 
The eigenfunction $\exp{(u_i)}$ corresponds to a part of 
Eq.~(2.5). 
Center and right panels: The results of mock data
analysis. The dotted line is an original SPF and the
solid line is reconstructed result by MEM. } 
\label{fig1}
\vspace{-0mm}
\end{figure}

An outstanding feature of Eq.(\ref{eq:spec_svd}) is that
it can be fitted to generic shape without restriction to
specific forms such as a sum of poles.
However, the resolution of course depends on the number of degree of
freedom in Eq.~(\ref{eq:spec_svd}), and also on $\omega$.
An example of eigenfunctions is displayed in the left panel of
Fig.~\ref{fig1}.
This figure indicates that the resolution of the function
becomes worse in large $\omega$ region, because the superposed
functions do not have enough variation.

This feature of the eigenfunction is also shown in the mock data 
analysis.
Center panel of Figure~\ref{fig1} shows the original (input) SPF and the
reconstructed SPF from the correlator which is constructed
by the original SPF with a random Gaussian noise.
The original SPF has three peaks of the same width and
hight at each $\omega$.
When the noise of correlator is not so small, the reconstructed SPF
does not agree with the original one;
there is a tendency that the peak becomes broader
than the original one at high energy region. 
The peak positions are correctly reproduced in this case.

The right panel of Figure~\ref{fig1} shows more interesting example.
When the width of a peak is narrow, MEM reproduces the shape
rather well.
However, for a case of large width, MEM fails to reproduce the
shape of the original peak.

\subsection{ Default model function }
\label{sect:def}

As mentioned in section \ref{sect:omem}, 
MEM needs a default model function to define the entropy term $Q$.
Since small difference between a trial SPF and default
model function makes the entropy term large, 
the default model function strongly affects the result of MEM
when the quality of data is not sufficient.
Therefore the default model function should include only reliable
information we know beforehand. 
If not so, there is a risk the result might be controlled by hand.

In the case of QCD, prior knowledge for SPFs is not so many,
e.g. positivity and perturbative behavior at high energy region.
In the case of point correlators, a natural choice of the default model
function is the asymptotic behavior of the meson
correlators at large $\omega$ in perturbation theory.
We should remember, however, that such an asymptotic behavior is not
observed in practical simulation because of the finite lattice cutoff.

The risk caused by lack of reliable default model function is reduced by
the quality of data.
In fact,
this perturbative form has been successfully applied to problems at zero
temperature \cite{nakahara99,yamazaki01}.
When we do not have good quality of data and reliable default model
function, we have to check, at least, a default model function dependence
to estimate an systematic uncertainty for the results.



\section{ Application to finite temperature lattice QCD }
\label{sect:ft}

Since a temporal lattice extent, $N_t$, is restricted to $1/Ta_t$ on
finite temperature lattices,   
it is usually difficult to keep good quality of data as compared with
zero temperature. 
Therefore we have to check the reliability of the results.

One of good checks is to apply the method to the zero temperature
data in the same condition as at finite temperature.
To extract the SPF at finite temperature, at least, we
should successfully reproduce the zero temperature SPF
from the zero temperature correlator but the number of
data points restricted to $1/Ta_t$.
We show these checks with our lattice data, which
was obtained on an anisotropic lattice with $\beta=6.10$ and
the renormalized anisotropy $\xi=4$
having the spatial cutoff
$a_{\sigma}^{-1}=2.030(13)$ GeV \cite{umeda02}.

\begin{figure}[tb]
\center{
\includegraphics[width=49mm]{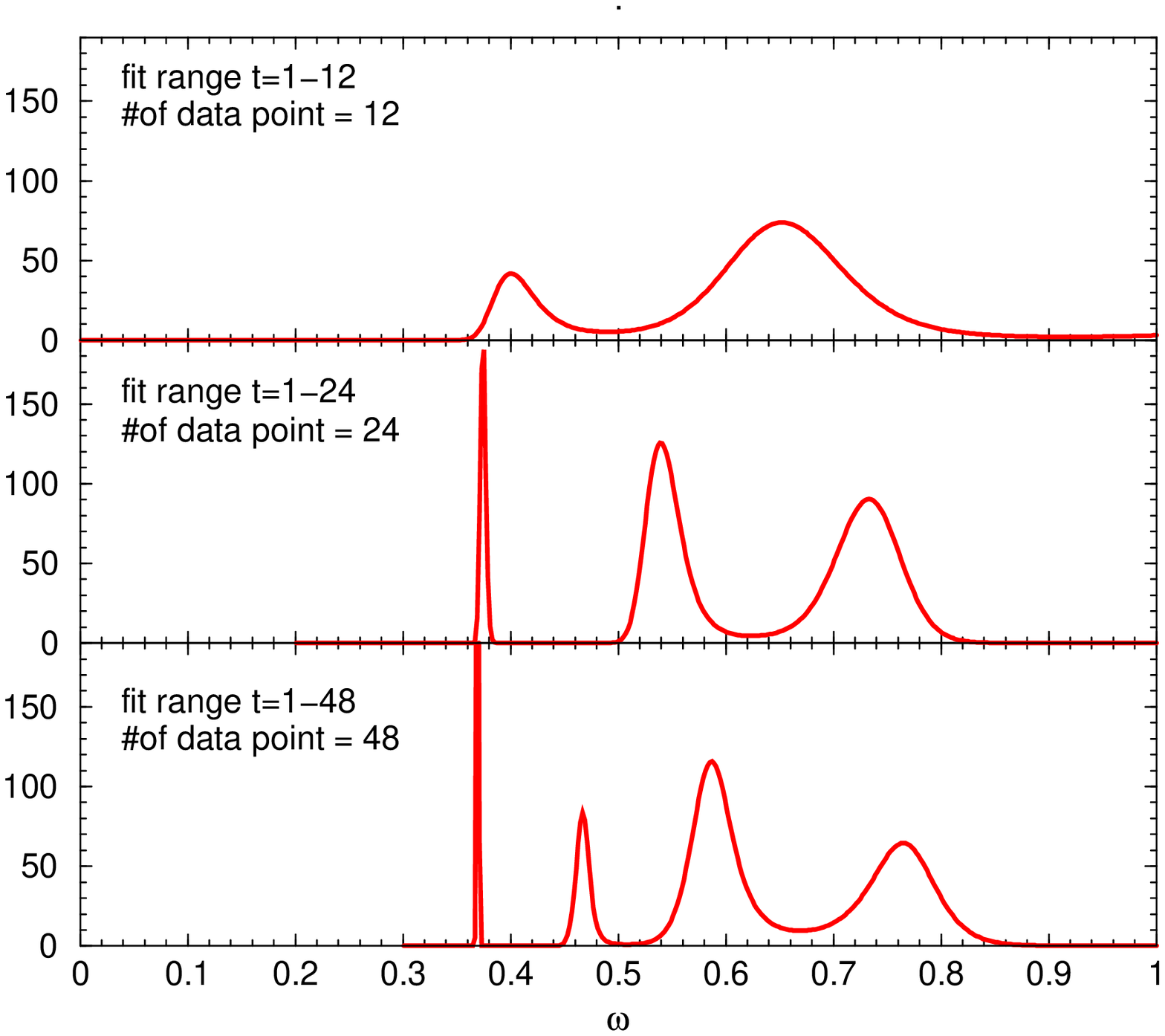}
\includegraphics[width=49mm]{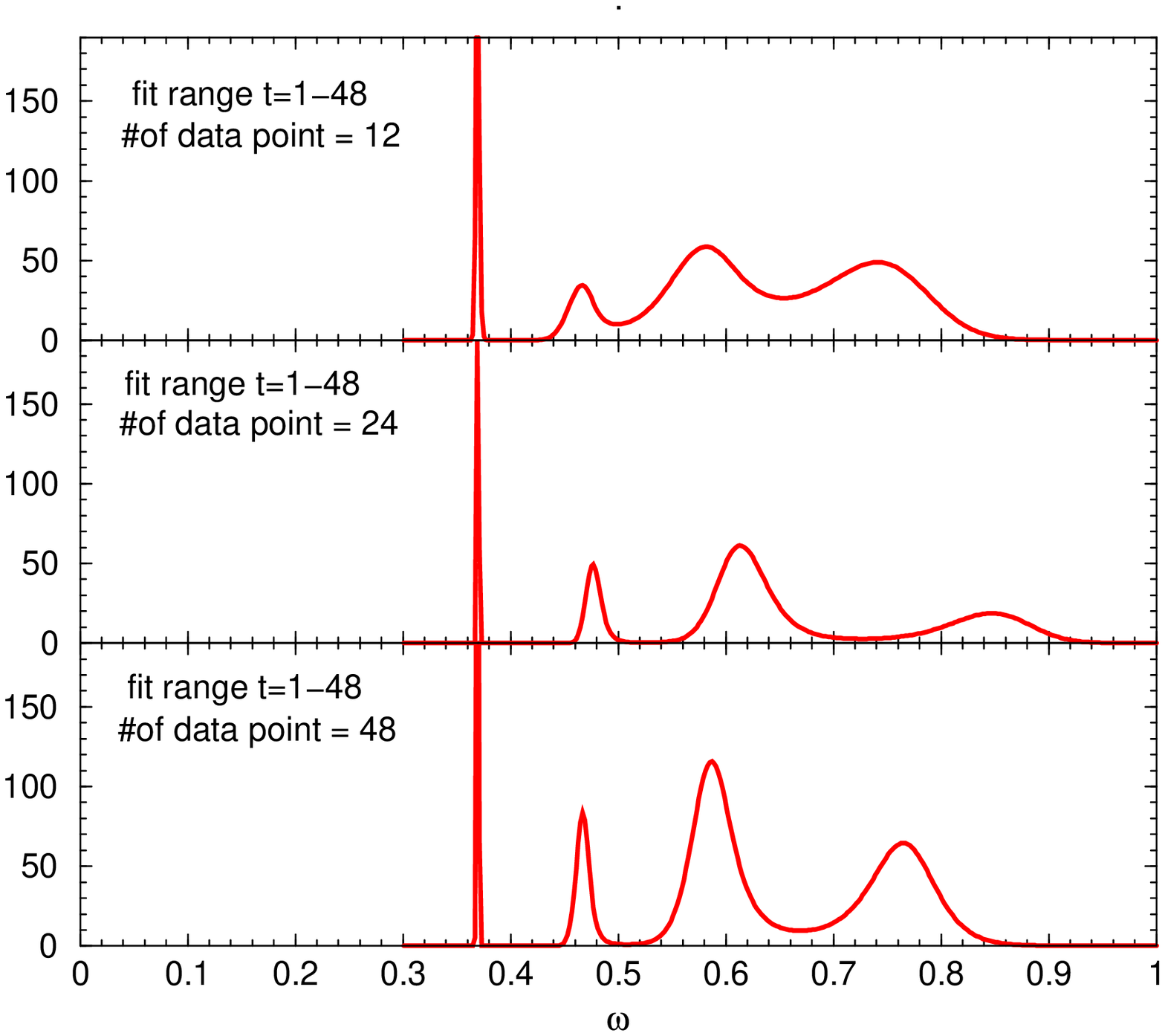}
\includegraphics[width=49mm]{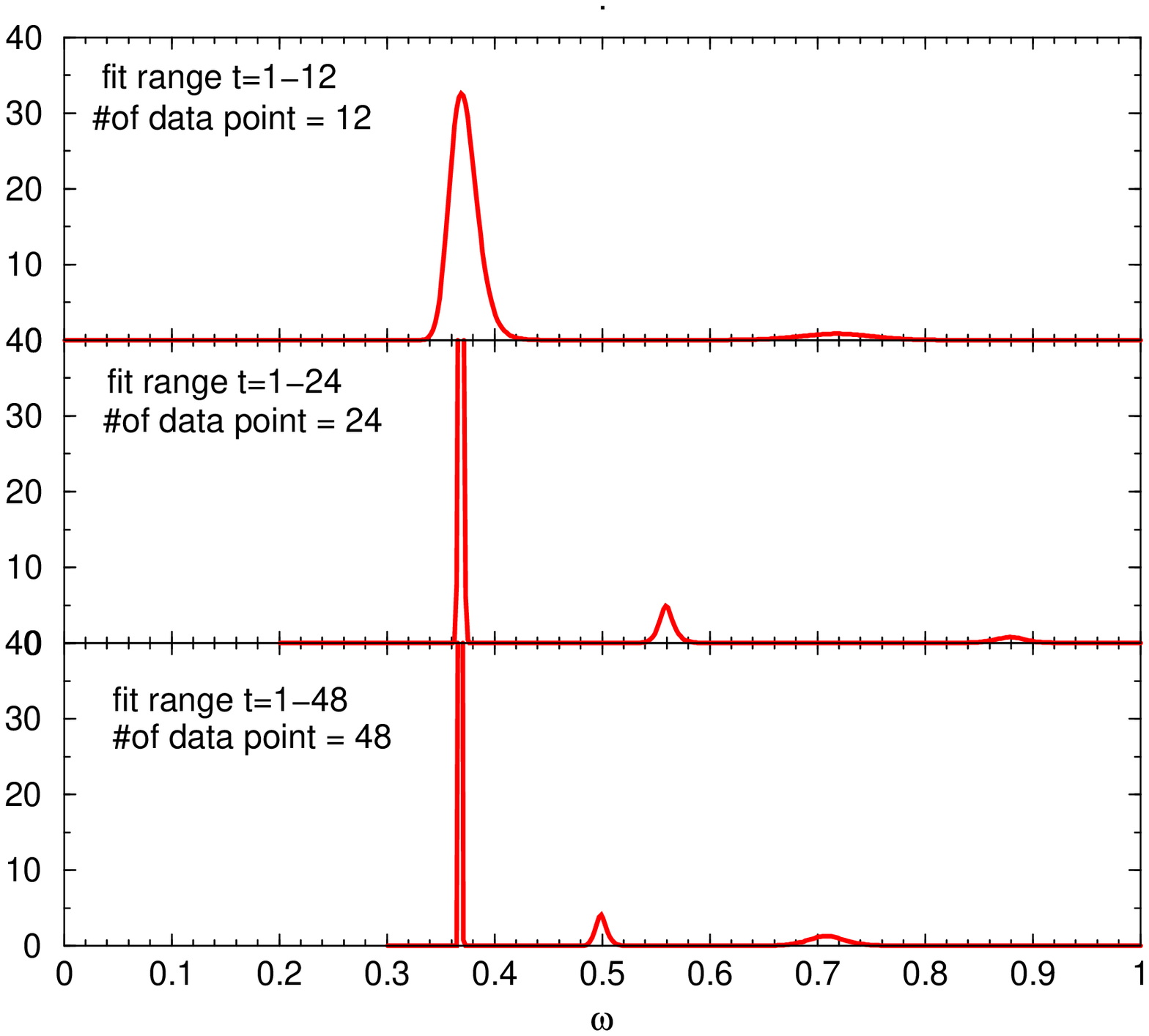}}
\vspace{-0mm}
\caption{Tests of MEM for restricting number of degree of freedom in
 the zero temperature correlator. The first two panels show the results
 with a point operator, and the last is with a smeared operator.}
\label{fig2}
\vspace{-0mm}
\end{figure}

We apply MEM to the correlator with restricted numbers of degree of freedom.
The results with two types of such restrictions are displayed in
Fig.~\ref{fig2}.
The left panel shows the dependence of the result on $t_{max}$, the
maximum $t$ of the correlator used in the analysis. This case
corresponds to the situation at $T>0$. MEM fails to reproduce even the
lowest peak for $t_{max}\le 16$. The center panel shows the results when
one alternatively skips several time slices in the analysis.
This case corresponds to the coarsening of the temporal lattice spacing.
Even for $t_{sep}=8$ for which the number of data point is 6, MEM at
least reproduces the correct lowest peak position while the resolution
is not enough. These result indicate that the physical region of the
correlator as well as the number of the degrees of freedom is important
for MEM to work correctly.
The required region of $C(t)$ in the above analysis is
$t_{max}>O(0.5fm)$, which is not fulfilled around $T\sim T_c$.
This situation may be improved by smeared operators. The left panel
shows the results of MEM for the correlator with smeared operator. 
It is stable under the above two kinds of restriction for $t_{max}$ of
interest; at least the lowest peak position is correctly reproduced.

\begin{figure}[tb]
\center{
\includegraphics[width=55mm]{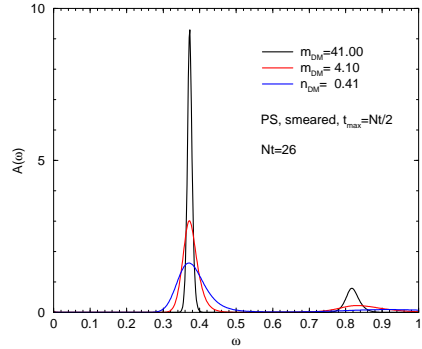}}
\vspace{-0mm}
\caption{The default model function dependence of SPF
for a correlator with the smeared operator at finite temperature.}
\label{fig3}
\vspace{-0mm}
\end{figure}

Next we show the default model function dependence for a correlator
with the smeared operator at finite temperature in Fig.~\ref{fig3}.
In this analysis we adopt the default model function of
$m(\omega)=m_{DM}\omega^2$, where $m_{DM}$ is determined by the 
perturbative asymptotic behavior of large $\omega$ region.
As mentioned in Sect.\ref{sect:def}, since the default model function is not
justified for lattice QCD simulation, we observe the default
model function dependence in order to estimate a systematic
uncertainty of the MEM results.
In Figure \ref{fig3} we change the $m_{DM}$ by factor 10, 1 and 0.1 respectively.
The peak position is stable while the width is rather sensitive to such
change of the parameter $m_{DM}$. 
These results indicate that it is difficult to discuss width of SPFs
quantitatively based on MEM analysis.

\section{ Conclusion }

In our previous paper \cite{umeda02} we have concluded the MEM is not
sufficient for 
quantitative study of the SPF from our lattice data even
if some smeared operators are adopted.
If we roughly know the shape of SPFs, standard
$\chi^2$ fit (or constrained curve fitting \cite{lepage02}) is rather
appropriate for quantitative studies.
Therefore we used MEM to find a rough image (fit-form) of SPFs and
performed $\chi^2$ fit (or constrained curve fitting) with 
this functional form, such as multi Breit-Wignar type function,
for more quantitative estimate of the width of the peak.
%
In conclusion, although MEM is powerful tool to extract the
SPFs from correlators, we have to use it carefully
taking its subtlety into account.

The simulation has been done on NEC SX-5 at
Research Center for Nuclear Physics, Osaka University
and Hitachi SR8000 at KEK (High Energy Accelerator Research
Organization).
T.U has been supported by the JSPS Research Fellowships
for Young Scientists.

\end{document}